\documentclass{iaus}
\usepackage{graphicx}

\title[Multiplicity in 5 $M_\odot$ Stars] 
{Multiplicity in 5 $M_\odot$ Stars}

\author[Nancy Remage Evans]   
{Nancy Remage Evans$^1$
 }

\affiliation{$^1$Smithsonian Astrophysical Observatory \\60 Garden
  St., MS 4, Cambridge MA 02138 USA \\ email: {\tt nevans@cfa.harvard.edu} \\}

\pubyear{2011}
\volume{272}  
\pagerange{}
\jname{Active OB Stars:Structure, Evolution, Mass Loss, and Critical Limits}
\editors{C. Neiner, G. Wade, G. Meynet, and G. Peters, eds.}
\begin{document}

\maketitle

\begin{abstract}

Binary/multiple status can affect stars at all stages of
their lifetimes: evolution onto the main sequence, 
properties on the main sequence, and subsequent evolution. 
5 $M_\odot$ stars have provided a wealth of information about the 
binary properties fairly massive stars.  
The combination of cool evolved primaries and
hot secondaries in Cepheids (geriatric B stars) have yielded
detailed information about the distribution of mass ratios.
and have also provided a surprisingly high fraction of
triple systems.  Ground-based radial velocity orbits combined with
satellite data from Hubble, FUSE, IUE, and Chandra are needed to
provide full information about the systems, including the
masses. As a recent example, X-ray observations can identify low
mass companions which are young enough to be physical
companions.  Typically binary status and properties (separation,
eccentricity, mass ratio) determine whether any  stage of 
evolution takes an exotic form.

\end{abstract}

\firstsection 
\section{Introduction}

Understanding the configurations involving massive stars, the
processes which shape them, and the objects they evolve into is
formidable--though being undertaken with equally formidable observing
and computing resources.  In the interplay of rotation, winds, and
magnetic fields binary/multiple status sometimes take on a leading
role, sometimes plays a relatively passive role.  It is, however, frequently a
significant factor and progress is being made in determining
binary/multiple properties.  This contribution discusses two aspects
for 5$M_\odot$  stars, the frequency of higher multiplicity systems and the
identification of low mass companions.  

{\bf Multiplicity} Cepheids (post-main sequence He burning stars of typically 5 $M_\odot$)
have provided some new insights into system multiplicity.  Frequently
they have hot companions which can be studied uncontaminated by the
light of the primary.  This has lead to the identification of triple
systems in a number of ways.  For the best studied sample (18 stars
with orbits and ultraviolet spectra of the companions), 44\% (possibly
50\%) are actually triple systems  (Evans 2005).

\begin{figure}[b]
\begin{center}
 \includegraphics[width=3.4in]{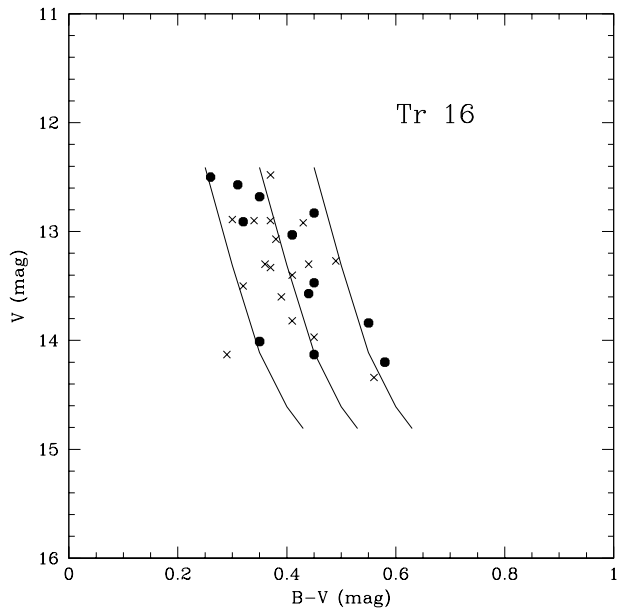}
 \caption{The sample of late B stars in Tr 16.  Lines show the ZAMS
   (center) with a range of E(B-V) = $\pm$ 0.1 mag.  Dots are X-ray
   sources; x's are not detected. }
   \label{fig1}
\end{center}
\end{figure}

{\bf Low Mass Companions} A second important area where we are obtaining new information about 
5$M_\odot$ systems is in the identification of low mass companions. These
are the most difficult companions to detect, either in photometric or
spectroscopic (radial velocity) studies.  We are exploring the use of
X-ray data to determine the fraction of late B (B3-A0) stars which
have low mass companions. Late B do not in general produce X-rays
themselves.  However, low mass stars (mid-F through K spectral types)
at the same  age ($\simeq$ 50 Myr) are strong X-ray producers,
much stronger than field stars of the same temperature.  Fig 1 shows
data from the Chandra ACIS image of Tr 16 (Evans, et al. 2011;  Townsley,
et al. 2011, Albacete-Colombo, et al. 2008).  With many exciting more massive
stars in Tr 16, little attention has been paid to late B stars, so the
first step was to establish a sample.  A list was compiled of stars
within 3' of $\eta$ Car (the center of Tr 16), with an appropriate
combination of V and B-V for the ZAMS (distance of 2.3 kpc,
E(B-V) = 0.55 mag $\pm$0.1 mag), and proper motions consistent
with cluster membership (Cudworth et al. 1993). 
 39\% of the stars are detected, spread through the luminosity range as
 would be expected for a random occurrence. These are identified as having low
 mass companions.  This approach is complementary to photometric or
interferometric surveys such as the O/B survey of  Mason et al. (2009)
and the IUE survey of Cepheids (Evans 1992) and probes new parameter space.

Funding for this work was provided by
Chandra X-ray Center NASA Contract NAS8-39073.




\end{document}